\documentclass{elsartL}
\usepackage{graphicx}
\usepackage{amssymb}
\usepackage{amsmath}
\usepackage{mathrsfs}
\usepackage{mathtools}
\usepackage{physics}
\usepackage{wasysym} % To display the astronomical sign of the Sun
\usepackage{mathtools}% http://ctan.org/pkg/mathtools
\newcommand{\avg}[1]{\left< #1 \right>} % For a nice display of the average value

\begin{document}

\begin{frontmatter}
\title{Analysis of luminosity measurements of the pre-white dwarf PG 1159-035: an approach featuring a dynamical database}
\author{E.~Matsinos{$^*$}}

\begin{abstract}In a previous work, those of the luminosity measurements of the pre-white dwarf PG 1159-035 which are available online yielded estimates for the optimal embedding dimension, for the dimensionality of the 
phase space reconstructed from these observations, and for the maximal Lyapunov exponent $\lambda$: the result $\lambda = (9.2 \pm 1.0 ({\rm stat.}) \pm 2.7 ({\rm syst.})) \cdot 10^{-2}~\Delta \tau^{-1}$ ($\Delta \tau=10$ 
s is the sampling interval in the measurements) was obtained, suggesting that the physical processes, underlying the variation of the luminosity of PG 1159-035, are chaotic.\\
\noindent An improved approach is employed in the present work in relation to the database of embedding vectors: instead of assigning each of the input time-series arrays either to the training or to the test set, the new 
approach features the creation of a dynamical database, i.e., of one which depends on the choice of the input test file. Although the size of the database is thus increased by a factor of about $2$ (compared to the previous 
study), the impact of this change on the important results is found to be insignificant. The estimate of this work for the maximal Lyapunov exponent ($\lambda = (8.9 \pm 0.7 ({\rm stat.}) \pm 1.9 ({\rm syst.})) \cdot 10^{-2}~\Delta \tau^{-1}$) 
is in very good agreement with the result of the earlier study.\\
\noindent {\it PACS 2010:} 05.10.-a; 05.45.-a; 05.45.Gg; 45.30.+s; 95.10.Fh
\end{abstract}
{$^*$}{Electronic mail: evangelos (dot) matsinos (at) sunrise (dot) ch}
\end{frontmatter}

\section{\label{sec:Introduction}Introduction}

Reported in Ref.~\cite{matsinos2018} were the results of an analysis of those of the luminosity measurements of the pre-white dwarf PG 1159-035 which had found their way into the `Time-series data source Archives: Santa 
F\'e Time Series Competition' database \cite{timeseries}. The seventeen time-series arrays of Ref.~\cite{timeseries} (to be called `original' from now on) appear to be snippets of the data acquired in the Whole Earth 
Telescope (WET) project in 1989. The rich power spectrum of the detected radiation was explored in Refs.~\cite{winget1991,costa2008}, resulting in the identification of $198$ pulsation modes of this celestial body. Also 
extracted in Refs.~\cite{winget1991,costa2008} was accurate information on the important physical properties of PG 1159-035, i.e., its rotation period, mass, magnetic field, and structure.

The original time-series arrays were band-passed in Ref.~\cite{matsinos2018}, using an elliptical filter and the band-pass corner frequencies of $1$ and $3$ mHz, which (according to Refs.~\cite{winget1991,costa2008}) delimit 
the region of interest in the power spectrum of PG 1159-035 as far as the non-radial gravity-wave ($g$-wave) pulsations of this celestial body are concerned. In Ref.~\cite{matsinos2018}, the filtered data was split into two 
parts of comparable sizes, one yielding the training (learning) set or database, the other the test set. The optimal embedding dimension $m_0$ was subsequently determined using Cao's method \cite{cao1997}: Cao's $E1$, 
representing the relative change of the average distance between neighbouring embedding vectors when increasing the embedding dimension by one unit, appeared to saturate in the vicinity of $10$, a result for $m_0$ which 
was corroborated by an analysis of the correlation dimension.

The extraction of the maximal Lyapunov exponent $\lambda$, characterising the rapidity of the exponential divergence between predictions and observations in chaotic systems, was subsequently pursued, by fitting a monotonic 
function to the out-of-sample prediction-error arrays $S (k)$. The original $S (k)$ arrays contain sizeable undulations, which hinder the establishment of the region in the ($k$,$S (k)$) plane within which a linear relationship 
(i.e., the signature of a chaotic dynamical system) holds. Parenthetically, a modification in the evaluation of the $S (k)$ arrays was put forward in Ref.~\cite{matsinos2018}, after introducing a weight which depends on the 
distance between two embedding vectors contributing to $S (k)$.

An estimate for the maximal Lyapunov exponent $\lambda$ was obtained for embedding dimensions $m$ between $10$ and $12$, within the domain of neighbourhood sizes $\epsilon$ for which a linear relationship with the correlation 
sums $C (\epsilon)$ (on a log-log plot) could be established: the final result $\lambda = (9.2 \pm 1.0 ({\rm stat.}) \pm 2.7 ({\rm syst.})) \cdot 10^{-2}~\Delta \tau^{-1}$ was obtained, where $\Delta \tau$ is the sampling 
interval in the measurements, namely $10$ s. Therefore, the findings of Ref.~\cite{matsinos2018} suggest that the physical processes, underlying the variation of the luminosity of PG 1159-035, are chaotic.

Two reasons call for further analysis of these measurements. The first one relates to the filtering of the original time-series arrays; this modification is expected to have a minor impact. The second reason relates to the 
creation of the database of embedding vectors; the significance of the impact of this modification on the important results of the study needs to be assessed.
\begin{itemize}
\item In Ref.~\cite{matsinos2018}, the filtering of the original time-series arrays was performed `left-to-right' (i.e., following the arrow of time). On the contrary, the original arrays will be subjected to a symmetrical, 
two-sided filtering in this work. This procedure has two advantages: first, it is expected to lead to the reduction of transient effects (which are due to the application of an IIR digital filter) at the start of each 
filtered array; second, it is expected to reduce the delay between the original and the filtered data. The elliptical filter of Ref.~\cite{matsinos2018} will be applied herein to the original time-series arrays, `left-to-right' 
and `right-to-left'. The output of the filtering procedure will be the average of the two filtered forms.
\item In Ref.~\cite{matsinos2018}, the training and test sets were fixed at the outset of the study: each of the (filtered) time-series arrays was assigned to one of the two sets, and remained in that set throughout the 
analysis. The procedure, put forward in Ref.~\cite{matsinos2018} for the determination of these two sets, led to the assignment of $12617$ measurements to the former set and of $14574$ to the latter. Despite the fact that 
the splitting of the data into (fixed) training and test sets is an efficient way in suppressing the effects of the temporal correlations\footnote{As the original time-series arrays of Ref.~\cite{timeseries} had obviously 
been obtained from a pool of measurements containing at least three times the amount of the chosen data, it is reasonable to assume that the data sets of Ref.~\cite{timeseries} are independent (i.e., no set contains elements 
which are temporally correlated with any of the elements of any other set).}, alternative analysis options, which are equally promising in terms of the suppression of these effects, are worth investigating (at least, for 
the sake of comparison and verification of the important results). One such possibility, featuring the creation of a dynamical database, is explored in this work (see Section \ref{sec:AnalysisFiltered}).
\end{itemize}
Apart from these two changes, the reader is addressed to Ref.~\cite{matsinos2018} for a concise description of the theoretical background upon which the study rests, as well as for other important details on the analysis.

\section{\label{sec:Filtering}The filtering of the time-series arrays of Ref.~\cite{timeseries}}

The original time-series arrays were submitted to the two-sided filtering procedure as described in the introduction. Some of the properties of the resulting filtered arrays are given in Table \ref{tab:FilteredData}. The 
values of the embedding delay $\nu$ come out consistent after the filtering, namely either $13$ or $14$ sampling intervals in all cases. Equally comforting is the assertion of the stationarity of the observations, as revealed 
by the p-values, which exceed the threshold of statistical significance $\mathrm{p}_{\rm min}$ assumed in this work (as in Ref.~\cite{matsinos2018}, $\mathrm{p}_{\rm min} = 1.00 \cdot 10^{-2}$).

\begin{table}%[h!]
{\bf \caption{\label{tab:FilteredData}}}Some properties of the filtered time-series arrays of PG 1159-035; $s$ stands for `signal' and represents the values of each array. The original measurements were subjected to a 
two-sided filtering, using the elliptical filter which had been applied to the data (`left-to-right') in Ref.~\cite{matsinos2018}.
\vspace{0.2cm}
\begin{center}
\begin{tabular}{|c|c|c|c|c|c|c|c|}
\hline
Data set & $s_{\rm min}$ & $s_{\rm max}$ & $s_{\rm max}-s_{\rm min}$ & $\avg{s}$ & rms & p-value & $\nu$\\
\hline
\hline
E01 & $-0.2232$ & $0.2314$ & $0.4546$ & $3.026 \cdot 10^{-4}$ & $1.083 \cdot 10^{-1}$ & $9.35 \cdot 10^{-1}$ & $14$\\
E02 & $-0.1823$ & $0.1774$ & $0.3596$ & $-4.645 \cdot 10^{-5}$ & $7.128 \cdot 10^{-2}$ & $9.26 \cdot 10^{-1}$ & $14$\\
E03 & $-0.1257$ & $0.1261$ & $0.2518$ & $1.018 \cdot 10^{-4}$ & $6.292 \cdot 10^{-2}$ & $9.28 \cdot 10^{-1}$ & $13$\\
E04 & $-0.1393$ & $0.1221$ & $0.2614$ & $1.210 \cdot 10^{-4}$ & $6.489 \cdot 10^{-2}$ & $6.90 \cdot 10^{-1}$ & $14$\\
E05 & $-0.0799$ & $0.0814$ & $0.1613$ & $-1.590 \cdot 10^{-4}$ & $3.763 \cdot 10^{-2}$ & $4.75 \cdot 10^{-1}$ & $13$\\
E06 & $-0.1446$ & $0.1469$ & $0.2914$ & $1.125 \cdot 10^{-4}$ & $6.022 \cdot 10^{-2}$ & $1.92 \cdot 10^{-1}$ & $13$\\
E07 & $-0.2075$ & $0.1980$ & $0.4055$ & $4.551 \cdot 10^{-5}$ & $8.024 \cdot 10^{-2}$ & $8.10 \cdot 10^{-2}$ & $14$\\
E08 & $-0.2733$ & $0.2739$ & $0.5473$ & $-7.094 \cdot 10^{-5}$ & $1.107 \cdot 10^{-1}$ & $9.60 \cdot 10^{-1}$ & $14$\\
E09 & $-0.2910$ & $0.2787$ & $0.5697$ & $5.784 \cdot 10^{-5}$ & $1.037 \cdot 10^{-1}$ & $6.12 \cdot 10^{-1}$ & $14$\\
E10 & $-0.2954$ & $0.2925$ & $0.5879$ & $6.431 \cdot 10^{-5}$ & $1.071 \cdot 10^{-1}$ & $8.22 \cdot 10^{-1}$ & $14$\\
E11 & $-0.1837$ & $0.1991$ & $0.3828$ & $1.190 \cdot 10^{-2}$ & $8.393 \cdot 10^{-2}$ & $4.03 \cdot 10^{-1}$ & $14$\\
E12 & $-0.2117$ & $0.2218$ & $0.4335$ & $6.363 \cdot 10^{-5}$ & $7.498 \cdot 10^{-2}$ & $6.92 \cdot 10^{-1}$ & $14$\\
E13 & $-0.2382$ & $0.2376$ & $0.4758$ & $-5.841 \cdot 10^{-7}$ & $8.442 \cdot 10^{-2}$ & $3.36 \cdot 10^{-1}$ & $14$\\
E14 & $-0.2485$ & $0.2339$ & $0.4824$ & $-1.653 \cdot 10^{-3}$ & $8.923 \cdot 10^{-2}$ & $7.19 \cdot 10^{-2}$ & $14$\\
E15 & $-0.2060$ & $0.1813$ & $0.3873$ & $-3.384 \cdot 10^{-4}$ & $8.481 \cdot 10^{-2}$ & $7.86 \cdot 10^{-1}$ & $14$\\
E16 & $-0.2079$ & $0.2044$ & $0.4123$ & $5.221 \cdot 10^{-3}$ & $8.084 \cdot 10^{-2}$ & $6.83 \cdot 10^{-1}$ & $14$\\
E17 & $-0.1373$ & $0.1339$ & $0.2712$ & $1.327 \cdot 10^{-4}$ & $5.457 \cdot 10^{-2}$ & $8.59 \cdot 10^{-1}$ & $13$\\
\hline
\end{tabular}
\end{center}
\vspace{0.5cm}
\end{table}

At this point, it needs to be mentioned that the direct comparison of the seventeen filtered arrays of Ref.~\cite{matsinos2018} and of the present study indicated that the impact of this modification on the important results 
would be insignificant.

\section{\label{sec:AnalysisFiltered}The analysis of the filtered time-series arrays}

Unlike in Ref.~\cite{matsinos2018}, all files will be allowed to enter the database in this work. This will be achieved as follows. The important results in a non-linear time-series analysis are obtained via iteration schemes 
over the input data. If the time-series arrays are split into training and test sets, one file is chosen to become the source of embedding vectors at each iteration step (let me call this data set `test file'). By suitably 
using the embedding vectors obtained from the test file along with those contained in the database, one extracts estimates for the quantities which are relevant in non-linear analyses, e.g., for the optimal embedding 
dimension, for the correlation dimension, for the maximal Lyapunov exponent, etc. As aforementioned, the database was fixed in Ref.~\cite{matsinos2018}. On the contrary, the database will be dynamically created in this work, 
comprising all available files, save the one chosen as test file at a given iteration step. The creation of such a dynamical database, depending on the choice of the input test file, enhances the statistics: it is expected to 
lead to a more reliable evaluation of the correlation sums $C (\epsilon)$ and of the out-of-sample prediction-error arrays $S (k)$, hence also of the estimates for the correlation dimension and for the maximal Lyapunov 
exponent.

Apart from the aforementioned modification, all details on the analysis of the filtered time-series arrays may be found in Ref.~\cite{matsinos2018}. The maximal embedding delay of Table \ref{tab:FilteredData}, namely $\nu = 14$, 
will be employed; the same value was also used in Ref.~\cite{matsinos2018}. The results of this work will be obtained exclusively with the $L^\infty$ distance: the distance between two vectors is defined as the maximal 
absolute difference between their corresponding components.

\subsection{\label{sec:Cao}Cao's method for the determination of the optimal embedding dimension}

\begin{figure}
\begin{center}
\includegraphics [width=15.5cm] {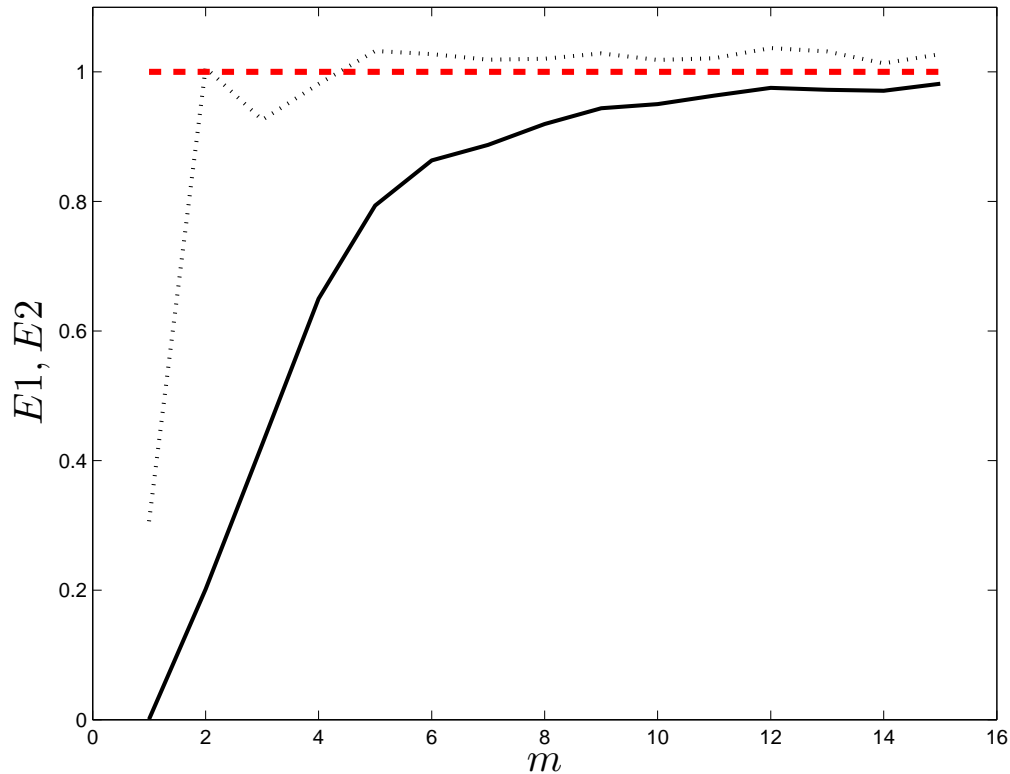}
\caption{\label{fig:CaoFromDatabase}Cao's $E1$ (straight line segments) and $E2$ (dotted line segments) for embedding dimensions $m$ up to $15$.}
\vspace{0.35cm}
\end{center}
\end{figure}

Cao's quantities $E1$ and $E2$ were evaluated from the filtered time-series arrays for embedding dimensions up to $m=15$, using (as the only input) $\nu=14$. The results are shown in Fig.~\ref{fig:CaoFromDatabase}.

\begin{figure}
\begin{center}
\includegraphics [width=15.5cm] {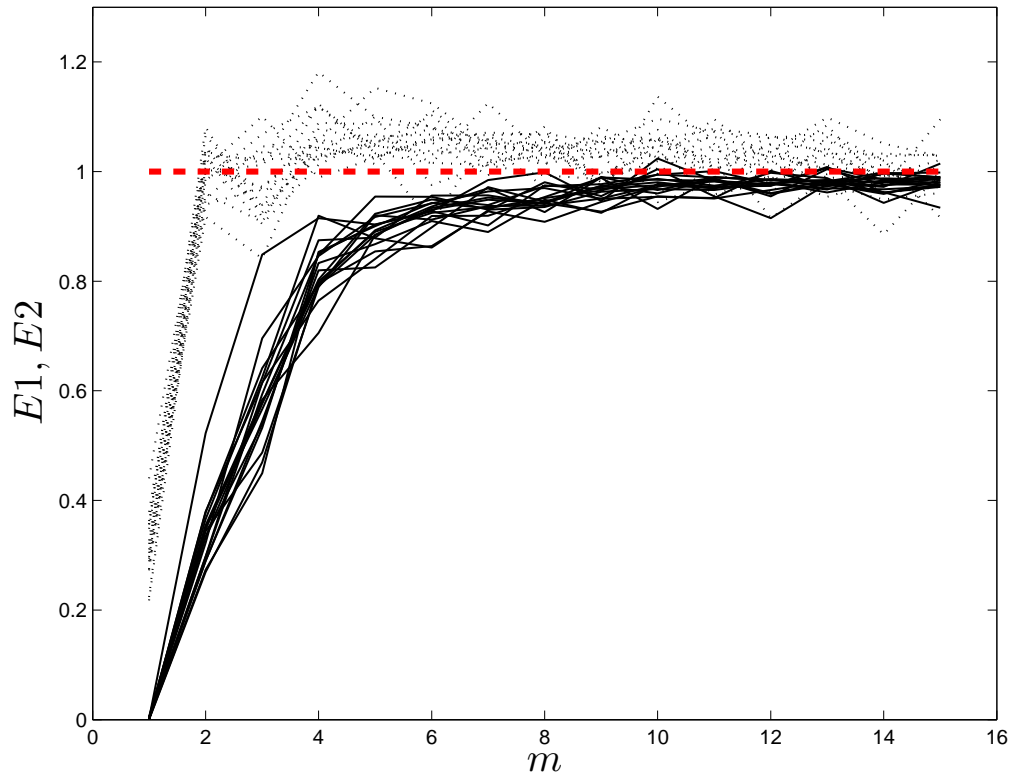}
\caption{\label{fig:Cao}Cao's $E1$ (straight line segments) and $E2$ (dotted line segments) for embedding dimensions $m$ up to $15$. These quantities were obtained from a separate analysis of the filtered time-series arrays. 
In order that their contributions to $E1$ and $E2$ be considered, the embedding vectors, involved in the evaluation, were required to have a temporal separation at least equal to $\nu \Delta \tau$.}
\vspace{0.35cm}
\end{center}
\end{figure}

The results of the separate analysis\footnote{To reduce the temporal correlations in the separate analysis of the data sets, all contributing (in the evaluation of $E1$ and $E2$) embedding vectors were required to have a 
temporal separation (constant distance between their corresponding elements) at least equal to the embedding delay $\nu=14$. A minimal temporal separation of $150~\Delta \tau$ (see Section \ref{sec:STSP}) was also imposed 
on the data, but yielded similar results.} of the filtered data sets are shown in Fig.~\ref{fig:Cao}. There is general agreement between Figs.~\ref{fig:CaoFromDatabase} and \ref{fig:Cao}, and they both also agree with the 
corresponding plots of Ref.~\cite{matsinos2018}. As in Ref.~\cite{matsinos2018}, $E1$ will be assumed to saturate in the vicinity of $10$; in other words, $m_0=10$.

\subsection{\label{sec:CorrelDim}Correlation dimension}

Introduced by Grassberger and Procaccia in 1983 \cite{grassberger1983}, the correlation dimension is a measure of the dimensionality of the phase space of the system under study. It is obtained from the correlation sums 
$C (\epsilon)$, which represent the frequentness of embedding vectors in the time series whose distance does not exceed a given neighbourhood size $\epsilon$. If no parts of the test files enter the database, the correlation 
sum $C (\epsilon)$ for embedding dimension $m$ is defined in Eq.~(8) of Ref.~\cite{matsinos2018}. One is interested in the domain of neighbourhood sizes for which the relationship between $\ln C (\epsilon)$ and $\ln \epsilon$ 
is linear. Within that region,
\begin{equation} \label{eq:EQ001}
\ln C (\epsilon) = \alpha \ln \epsilon + \beta \, \, \, ,
\end{equation}
where the slope $\alpha$ is identified with the quantity $\alpha (N,\epsilon)$ of Eq.~(7) of Ref.~\cite{matsinos2018}. At fixed $m$, the linearity between $\ln C (\epsilon)$ and $\ln \epsilon$ is investigated as in 
Ref.~\cite{matsinos2018}: starting from the original ($\ln \epsilon$,$\ln C (\epsilon)$) points, the point with the largest $\epsilon$ value (one point per iteration) was removed, until the resulting p-value (obtained from 
the $\chi^2$ value and the number of degrees of freedom in the linear fit, see Ref.~\cite{matsinos2018} for details) exceeded $\mathrm{p}_{\rm min}$, the threshold of statistical significance. Evidently, the surviving 
($\ln \epsilon$,$\ln C (\epsilon)$) points are those for which the linearity between $\ln C (\epsilon)$ and $\ln \epsilon$ is accepted at the assumed significance level.

To avoid the inclusion of noisy data in Ref.~\cite{matsinos2018}, a lower limit in the number of contributions was set in the evaluation of the correlation sums $C (\epsilon)$: points with fewer than $N_c=10$ contributions 
to $C (\epsilon)$ were not considered. To investigate the stability of the results, it was decided to make use of four $N_c$ values in this work: $10$, $20$, $50$, and $100$. The results for the extent of the linearity 
region and for the parameters of the linear fit, for these four $N_c$ choices, are contained in Table \ref{tab:DimensionLinf}.

\begin{table}%[h!]
{\bf \caption{\label{tab:DimensionLinf}}}Results of the linear fits of Eq.~(\ref{eq:EQ001}) for embedding dimensions between $3$ and $12$. The given domain $[ \epsilon_{\rm min}, \epsilon_{\rm max} ]$ corresponds to the 
$\epsilon$ domain within which the linearity between $\ln C (\epsilon)$ and $\ln \epsilon$ is accepted ($\mathrm{p} \geq \mathrm{p}_{\rm min}$). The blocks correspond to four choices of $N_c$, i.e., of the lower number of 
neighbours for acceptable contributions in the evaluation of the correlation sums $C (\epsilon)$.
\vspace{0.2cm}
\begin{center}
\begin{tabular}{|c|c|c|c|c|c|c|}
\hline
$m$ & $\epsilon_{\rm min}$ & $\epsilon_{\rm max}$ & $\alpha$ & $\delta \alpha$ & $\beta$ & $\delta \beta$\\
\hline
\hline
\multicolumn{7}{|c|}{$N_c = 10$} \\
\hline
$3$ & $6.00 \cdot 10^{-3}$ & $1.20 \cdot 10^{-2}$ & $2.9409$ & $0.0055$ & $6.612$ & $0.025$\\
$4$ & $6.00 \cdot 10^{-3}$ & $1.10 \cdot 10^{-2}$ & $3.900$ & $0.018$ & $9.561$ & $0.083$\\
$5$ & $6.00 \cdot 10^{-3}$ & $1.10 \cdot 10^{-2}$ & $4.828$ & $0.040$ & $12.50$ & $0.18$\\
$6$ & $6.00 \cdot 10^{-3}$ & $1.10 \cdot 10^{-2}$ & $5.893$ & $0.081$ & $16.08$ & $0.37$\\
$7$ & $6.00 \cdot 10^{-3}$ & $1.00 \cdot 10^{-2}$ & $7.41$ & $0.32$ & $21.9$ & $1.5$\\
$8$ & $7.00 \cdot 10^{-3}$ & $1.80 \cdot 10^{-2}$ & $6.645$ & $0.045$ & $17.06$ & $0.18$\\
$9$ & $8.00 \cdot 10^{-3}$ & $1.80 \cdot 10^{-2}$ & $7.479$ & $0.089$ & $19.58$ & $0.37$\\
$10$ & $1.00 \cdot 10^{-2}$ & $1.70 \cdot 10^{-2}$ & $9.09$ & $0.26$ & $25.3$ & $1.1$\\
$11$ & $1.10 \cdot 10^{-2}$ & $1.70 \cdot 10^{-2}$ & $10.10$ & $0.51$ & $28.5$ & $2.1$\\
$12$ & $1.40 \cdot 10^{-2}$ & $2.60 \cdot 10^{-2}$ & $8.661$ & $0.072$ & $21.69$ & $0.27$\\
\hline
\multicolumn{7}{|c|}{$N_c = 20$} \\
\hline
$3$ & $6.00 \cdot 10^{-3}$ & $1.20 \cdot 10^{-2}$ & $2.9409$ & $0.0055$ & $6.612$ & $0.025$\\
$4$ & $6.00 \cdot 10^{-3}$ & $1.10 \cdot 10^{-2}$ & $3.900$ & $0.018$ & $9.561$ & $0.083$\\
$5$ & $6.00 \cdot 10^{-3}$ & $1.10 \cdot 10^{-2}$ & $4.828$ & $0.040$ & $12.50$ & $0.18$\\
$6$ & $6.00 \cdot 10^{-3}$ & $1.10 \cdot 10^{-2}$ & $5.893$ & $0.081$ & $16.08$ & $0.37$\\
$7$ & $7.00 \cdot 10^{-3}$ & $1.10 \cdot 10^{-2}$ & $6.89$ & $0.18$ & $19.42$ & $0.82$\\
$8$ & $7.00 \cdot 10^{-3}$ & $1.80 \cdot 10^{-2}$ & $6.645$ & $0.045$ & $17.06$ & $0.18$\\
$9$ & $8.00 \cdot 10^{-3}$ & $1.80 \cdot 10^{-2}$ & $7.479$ & $0.089$ & $19.58$ & $0.37$\\
$10$ & $1.10 \cdot 10^{-2}$ & $1.70 \cdot 10^{-2}$ & $9.02$ & $0.26$ & $25.0$ & $1.1$\\
$11$ & $1.30 \cdot 10^{-2}$ & $1.60 \cdot 10^{-2}$ & $10.8$ & $1.1$ & $31.6$ & $4.5$\\
$12$ & $1.40 \cdot 10^{-2}$ & $2.60 \cdot 10^{-2}$ & $8.661$ & $0.072$ & $21.69$ & $0.27$\\
\hline
\end{tabular}
\end{center}
\vspace{0.5cm}
\end{table}

\newpage
\begin{table*}
{\bf Table \ref{tab:DimensionLinf} continued}
\vspace{0.2cm}
\begin{center}
\begin{tabular}{|c|c|c|c|c|c|c|}
\hline
$m$ & $\epsilon_{\rm min}$ & $\epsilon_{\rm max}$ & $\alpha$ & $\delta \alpha$ & $\beta$ & $\delta \beta$\\
\hline
\multicolumn{7}{|c|}{$N_c = 50$} \\
\hline
$3$ & $6.00 \cdot 10^{-3}$ & $1.20 \cdot 10^{-2}$ & $2.9409$ & $0.0055$ & $6.612$ & $0.025$\\
$4$ & $6.00 \cdot 10^{-3}$ & $1.10 \cdot 10^{-2}$ & $3.900$ & $0.018$ & $9.561$ & $0.083$\\
$5$ & $6.00 \cdot 10^{-3}$ & $1.10 \cdot 10^{-2}$ & $4.828$ & $0.040$ & $12.50$ & $0.18$\\
$6$ & $6.00 \cdot 10^{-3}$ & $1.10 \cdot 10^{-2}$ & $5.893$ & $0.081$ & $16.08$ & $0.37$\\
$7$ & $7.00 \cdot 10^{-3}$ & $1.10 \cdot 10^{-2}$ & $6.89$ & $0.18$ & $19.42$ & $0.82$\\
$8$ & $8.00 \cdot 10^{-3}$ & $1.80 \cdot 10^{-2}$ & $6.638$ & $0.045$ & $17.03$ & $0.19$\\
$9$ & $1.00 \cdot 10^{-2}$ & $1.80 \cdot 10^{-2}$ & $7.464$ & $0.095$ & $19.52$ & $0.39$\\
$10$ & $1.20 \cdot 10^{-2}$ & $1.70 \cdot 10^{-2}$ & $8.95$ & $0.30$ & $24.7$ & $1.2$\\
$11$ & $1.40 \cdot 10^{-2}$ & $2.30 \cdot 10^{-2}$ & $8.285$ & $0.087$ & $21.01$ & $0.34$\\
$12$ & $1.50 \cdot 10^{-2}$ & $2.60 \cdot 10^{-2}$ & $8.640$ & $0.068$ & $21.61$ & $0.25$\\
\hline
\multicolumn{7}{|c|}{$N_c = 100$} \\
\hline
$3$ & $6.00 \cdot 10^{-3}$ & $1.20 \cdot 10^{-2}$ & $2.9409$ & $0.0055$ & $6.612$ & $0.025$\\
$4$ & $6.00 \cdot 10^{-3}$ & $1.10 \cdot 10^{-2}$ & $3.900$ & $0.018$ & $9.561$ & $0.083$\\
$5$ & $6.00 \cdot 10^{-3}$ & $1.10 \cdot 10^{-2}$ & $4.828$ & $0.040$ & $12.50$ & $0.18$\\
$6$ & $6.00 \cdot 10^{-3}$ & $1.10 \cdot 10^{-2}$ & $5.893$ & $0.081$ & $16.08$ & $0.37$\\
$7$ & $8.00 \cdot 10^{-3}$ & $1.20 \cdot 10^{-2}$ & $6.41$ & $0.18$ & $17.19$ & $0.81$\\
$8$ & $9.00 \cdot 10^{-3}$ & $1.80 \cdot 10^{-2}$ & $6.635$ & $0.049$ & $17.02$ & $0.20$\\
$9$ & $1.00 \cdot 10^{-2}$ & $1.80 \cdot 10^{-2}$ & $7.464$ & $0.095$ & $19.52$ & $0.39$\\
$10$ & $1.30 \cdot 10^{-2}$ & $1.70 \cdot 10^{-2}$ & $8.84$ & $0.36$ & $24.2$ & $1.5$\\
$11$ & $1.40 \cdot 10^{-2}$ & $2.30 \cdot 10^{-2}$ & $8.285$ & $0.087$ & $21.01$ & $0.34$\\
$12$ & $1.50 \cdot 10^{-2}$ & $2.60 \cdot 10^{-2}$ & $8.640$ & $0.068$ & $21.61$ & $0.25$\\
\hline
\end{tabular}
\end{center}
\vspace{0.5cm}
\end{table*}

From the entries of Table \ref{tab:DimensionLinf}, one concludes that the dependence of the important results on $N_c$ is weak; this is not a surprise, as the $N_c$ cut affects only the points with the lowest $\epsilon$ 
values, hence those of the points which are accompanied by the largest statistical uncertainty. Using only sufficient embeddings ($10 \leq m \leq 12$), one obtains for the slope $\alpha$ the values of $8.72 \pm 0.15$, 
$8.69 \pm 0.12$, $8.52 \pm 0.13$, and $8.51 \pm 0.12$ for $N_c = 10$, $20$, $50$, and $100$, respectively. The estimate for the slope $\alpha$, obtained in Ref.~\cite{matsinos2018} for $N_c=10$, was $8.830 \pm 0.062$. One 
word about the increased uncertainties is due. In the approach followed in this work, the enhancement of statistics in the evaluation of the correlation sums $C (\epsilon)$ leads to smaller uncertainties $\delta \ln C (\epsilon)$, 
thus imposing more stringent conditions in the linear fits and, as it so happens, restricting the $\epsilon$ domain within which the linearity between $\ln C (\epsilon)$ and $\ln \epsilon$ is fulfilled. As a result, the 
estimates for the parameters of the linear fit are generally accompanied herein by larger fitted uncertainties (compared to those of Ref.~\cite{matsinos2018}).

\begin{figure}
\begin{center}
\includegraphics [width=15.5cm] {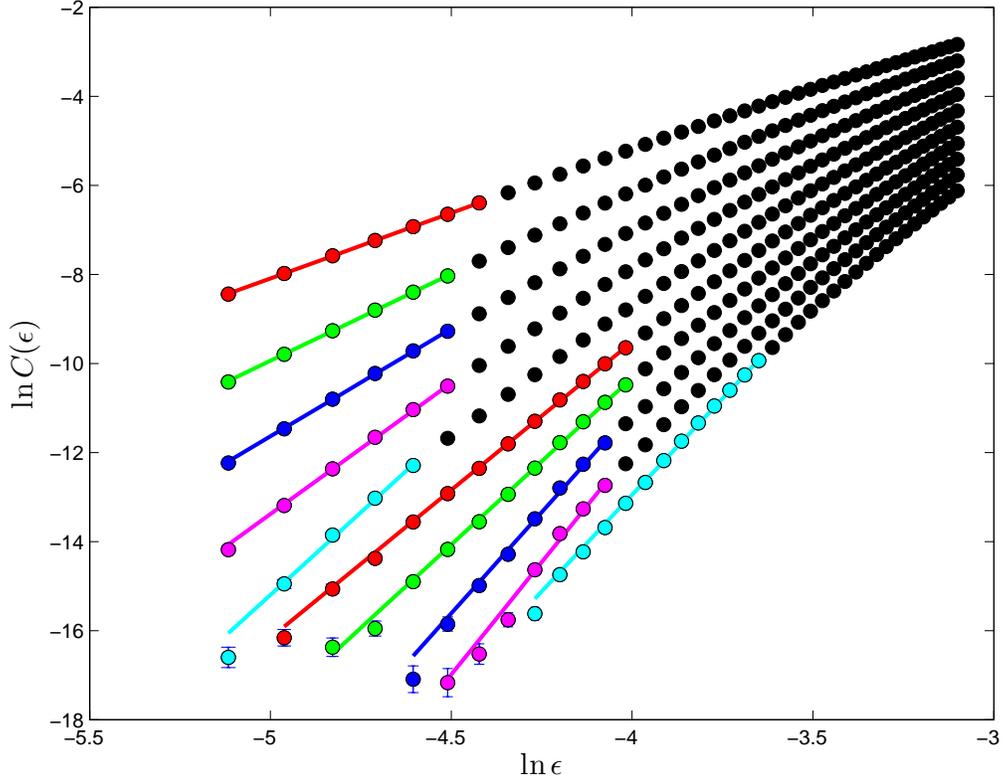}
\caption{\label{fig:CorrelationSums}The ($\ln \epsilon$,$\ln C (\epsilon)$) scatter plots for embedding dimensions $m=3$ (top) to $m=12$ (bottom). Weighted least-squares fits were performed on the data (see Table 
\ref{tab:DimensionLinf}), separately for each embedding dimension, in the $\epsilon$ domain within which the linearity between $\ln C (\epsilon)$ and $\ln \epsilon$ holds (points and straight lines in colour); the data 
outside these $\epsilon$ domains are also shown (in black).}
\vspace{0.35cm}
\end{center}
\end{figure}

\begin{figure}
\begin{center}
\includegraphics [width=15.5cm] {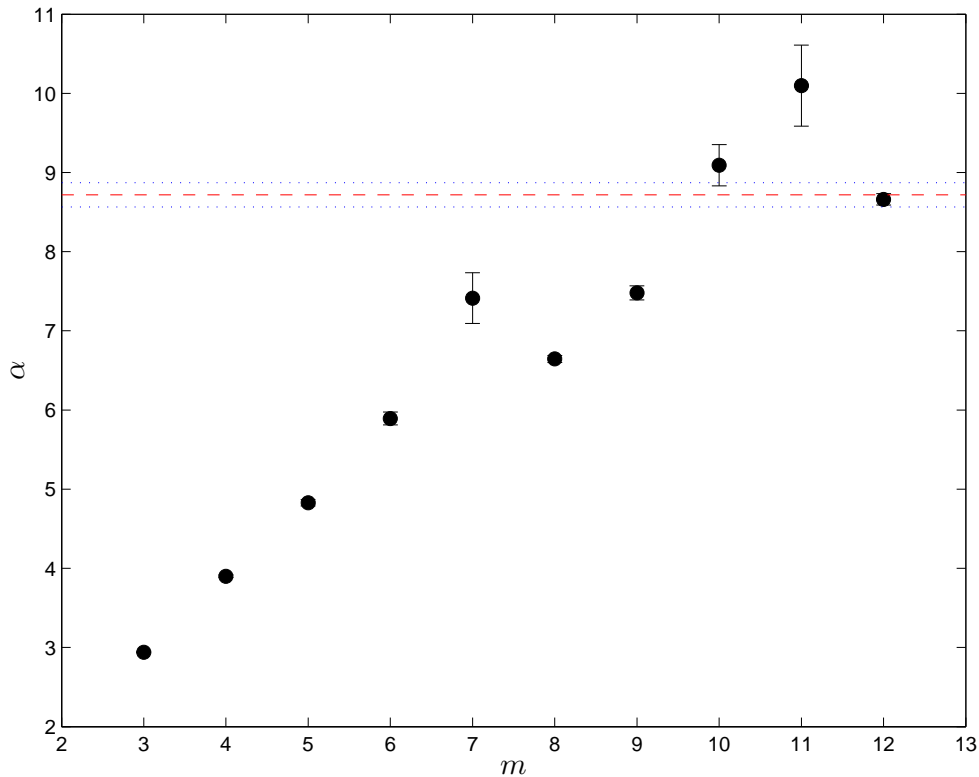}
\caption{\label{fig:CorrelationDimension}The values of the slope $\alpha$ of the linear fits of Eq.~(\ref{eq:EQ001}). The red dashed line represents the weighted average over the $\alpha$ values for sufficient embeddings 
($10 \leq m \leq 12$), see Table \ref{tab:DimensionLinf}, block corresponding to $N_c=10$. The blue dotted lines represent the $1 \sigma$ limits of the statistical uncertainty, corrected for the quality of the reproduction 
of the three input $\alpha$ values by their weighted average. The slightly misplaced $\alpha$ value for $m=11$ moves towards its two neighbours when larger $N_c$ cuts are used, see Table \ref{tab:DimensionLinf}.}
\vspace{0.35cm}
\end{center}
\end{figure}

The estimates for the slope $\alpha$ agree within the uncertainties, and one does not have good reasons to depart from the choice $N_c=10$ of Ref.~\cite{matsinos2018}. Figures \ref{fig:CorrelationSums} and \ref{fig:CorrelationDimension} 
contain the main results regarding the correlation dimension (for $N_c=10$).

Before addressing the extraction of the maximal Lyapunov exponent, one word of caution is due. In this work, the contributions to the correlation sums $C (\epsilon)$ from two neighbouring vectors - belonging to two data sets, 
say, $i$ and $j$ - are bound to appear twice, once when the data set $i$ is chosen to be the input test file and a second time when the data set $j$ is selected. Provided that the quantities $C (\epsilon)$ are normalised, 
one might think that this `double counting' is inessential. In reality, the $C (\epsilon)$ values do remain unchanged when using the complete database for each chosen test file (i.e., the embedding vectors constructed from 
all files, save the one chosen as test file at that iteration step), yet their uncertainties (which are required in the test of linearity between $\ln C (\epsilon)$ and $\ln \epsilon$) are obviously affected. Therefore, when 
the iteration is made over all data sets and the database comprises all other files, one must bear in mind to take this `double counting' into account in order to estimate $\delta \ln C (\epsilon)$ correctly. As a matter of 
fact, the implementation may be made in a clever way, avoiding this pitfall altogether; for instance, one may assume the order of Table \ref{tab:FilteredData} and allow only the data sets \emph{after} the chosen test file 
to become part of the database corresponding to that input file.

\subsection{\label{sec:Lyapunov}Maximal Lyapunov exponent}

The estimates for the maximal Lyapunov exponent were obtained as described in Section 5.6 of Ref.~\cite{matsinos2018}. The out-of-sample prediction-error arrays $S (k)$ were evaluated using the appropriate variation of $m$ 
and $\epsilon$ (i.e., within the linearity region in the ($\ln \epsilon$,$\ln C (\epsilon)$) scatter plots) and the distance-dependent weights $w_{ij} = 1 - (d_{ij}/\epsilon)^2$. Subsequently, their undulations were removed 
by fitting the monotonic function
\begin{equation} \label{eq:EQ002}
S (k) = \ln \left[ x_1 \exp\left[ x_2 \left( 1 + \frac{x_3}{x_1} \right) k \right] + x_3 \right] \, \, \, ,
\end{equation}
where the parameters $x_{1,2,3}$ are associated with the variation of $S (k)$ between $k=0$ and saturation, the maximal Lyapunov exponent ($\lambda$), and the saturation level of $S (k)$, respectively; the expansion of $S (k)$ 
of Eq.~(\ref{eq:EQ002}) for small $k$ values is: $S (k) \approx \ln (x_1 + x_3) + x_2 k$. The MINUIT package \cite{jms} of the CERN library (FORTRAN version) was used in the optimisation.

\begin{figure}
\begin{center}
\includegraphics [width=15.5cm] {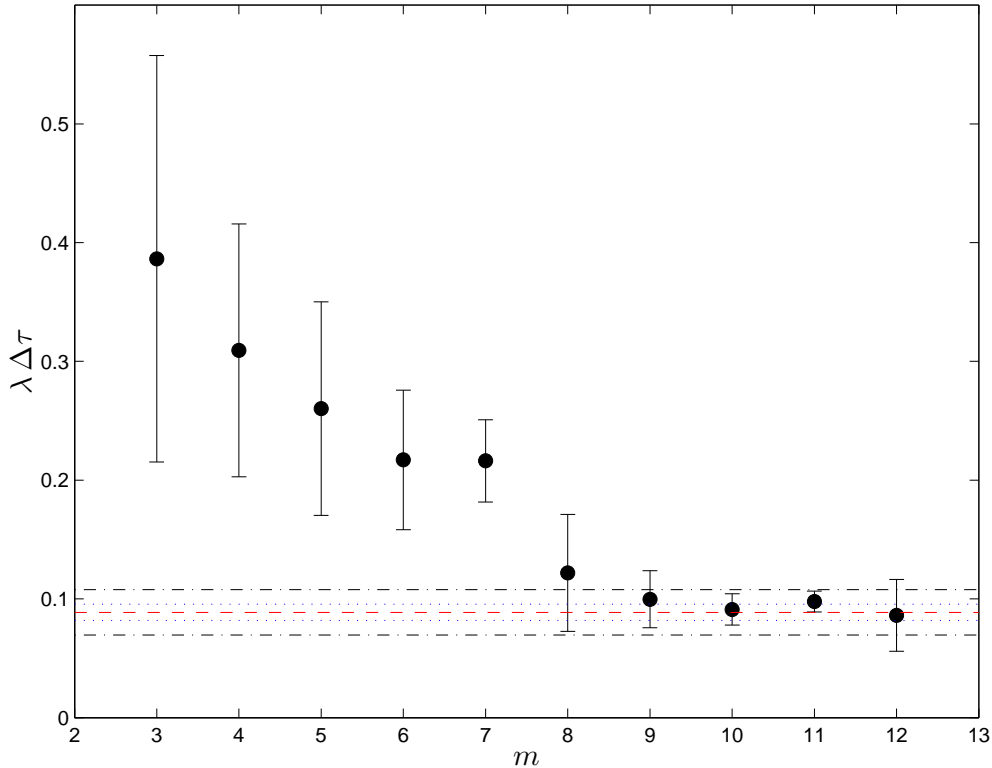}
\caption{\label{fig:Lyapunov}The maximal Lyapunov exponents extracted from the filtered luminosity measurements of PG 1159-035 using Eq.~(\ref{eq:EQ002}). The sum of the statistical and systematic uncertainties is shown for 
each data point; the statistical uncertainties have been corrected for the quality of each fit. The red dashed line represents the grand mean over the $\epsilon$ values for sufficient embeddings ($10 \leq m \leq 12$). The 
blue dotted lines represent the $1 \sigma$ limits of the statistical uncertainty of the grand mean, whereas the black dash-dotted lines correspond to the ($1 \sigma$ limits of the) systematic uncertainty.}
\vspace{0.35cm}
\end{center}
\end{figure}

As in Ref.~\cite{matsinos2018}, the results of the fits with unreasonably large $\chi^2$ values were removed after the application of a $\chi^2$ cut, set equal to twice the median value of the original $\chi^2$ distribution; 
$62$ (out of the $81$ original) fits with $\chi^2 \lesssim 264.23$ were accepted and yielded the maximal Lyapunov exponents shown in Fig.~\ref{fig:Lyapunov}. The values, obtained for sufficient embeddings ($10 \leq m \leq 12$), 
were found compatible: the grand-mean value $\lambda = (8.9 \pm 0.7 ({\rm stat.}) \pm 1.9 ({\rm syst.})) \cdot 10^{-2}~\Delta \tau^{-1}$ of this work is in very good agreement with the (slightly less precise) result of 
Ref.~\cite{matsinos2018}: $\lambda = (9.2 \pm 1.0 ({\rm stat.}) \pm 2.7 ({\rm syst.})) \cdot 10^{-2}~\Delta \tau^{-1}$. It should be reminded that the first uncertainty is statistical (average over fitted uncertainties, 
corrected for the quality of each fit), whereas the second one is systematic (reflecting the variation of $\lambda$ with $\epsilon$ for the sufficient embeddings).

\subsection{\label{sec:STSP}Space-time separation plot}

I decided to include in this paper a few words on the widely-used space-time separation plot (STSP), whose visual inspection enables the extraction of an estimate for the temporal interval within which the elements of a 
given time series are (temporally) correlated. It has been known since a long time (e.g., see Ref.~\cite{kantz1997} and the references therein) that the temporal correlations affect the evaluation of the correlation sums 
$C (\epsilon)$, hence the determination of the estimate for the correlation dimension.

Despite the fact that no use of the STSP is called for herein, I will nevertheless concisely describe the method for the sake of completeness, and obtain results from the available data of PG 1159-035. Such information may 
be useful to those interested in including more of the available data in the dynamically-created database (and therefore need to fix $n_{\rm min}$ in Eq.~(6) of Ref.~\cite{matsinos2018}).

For a time-series array comprised of independent elements, the probability $P (d_{ij} \leq \epsilon)$ (i.e., the probability that the distance $d_{ij}$ between two embedding vectors $i$ and $j$ does not exceed a neighbourhood 
size $\epsilon$) is expected to be only $\epsilon$-dependent; it should not depend on the temporal separation $\Delta t$ between the embedding vectors $i$ and $j$. Therefore, if the percentiles of the $d_{ij}$ distribution 
are plotted in $2$ dimensions, i.e., as functions of $\epsilon$ (vertical axis) and of $\Delta t$ (horizontal axis), the resulting curves should (in the absence of temporal correlations) be horizontal lines. On the contrary, 
if temporal correlations are present, a dependence of the percentiles of the $d_{ij}$ distribution on $\Delta t$ is unavoidable. To obtain an estimate for the minimal temporal separation between two elements of a time series, 
so that these elements be considered uncorrelated, Provenzale and collaborators \cite{provenzale1992} put forward the following procedure:
\begin{itemize}
\item fixation of $\Delta t$ at several values,
\item extraction of the $d_{ij}$ distribution for embedding vectors separated by the chosen $\Delta t$ value, and
\item determination of the $\epsilon$ values corresponding to certain percentiles of the $d_{ij}$ distribution (e.g., $10 \%$, $20 \%$, etc.) at the chosen $\Delta t$ value.
\end{itemize}
The choice of the temporal interval which enables the mitigation/suppression of the effects of the temporal correlations simply reduces to the extraction (from the STSP) of the time above which the percentiles of the $d_{ij}$ 
distribution are not dependent (at least, in a discernible way) on $\Delta t$.

\begin{figure}
\begin{center}
\includegraphics [width=15.5cm] {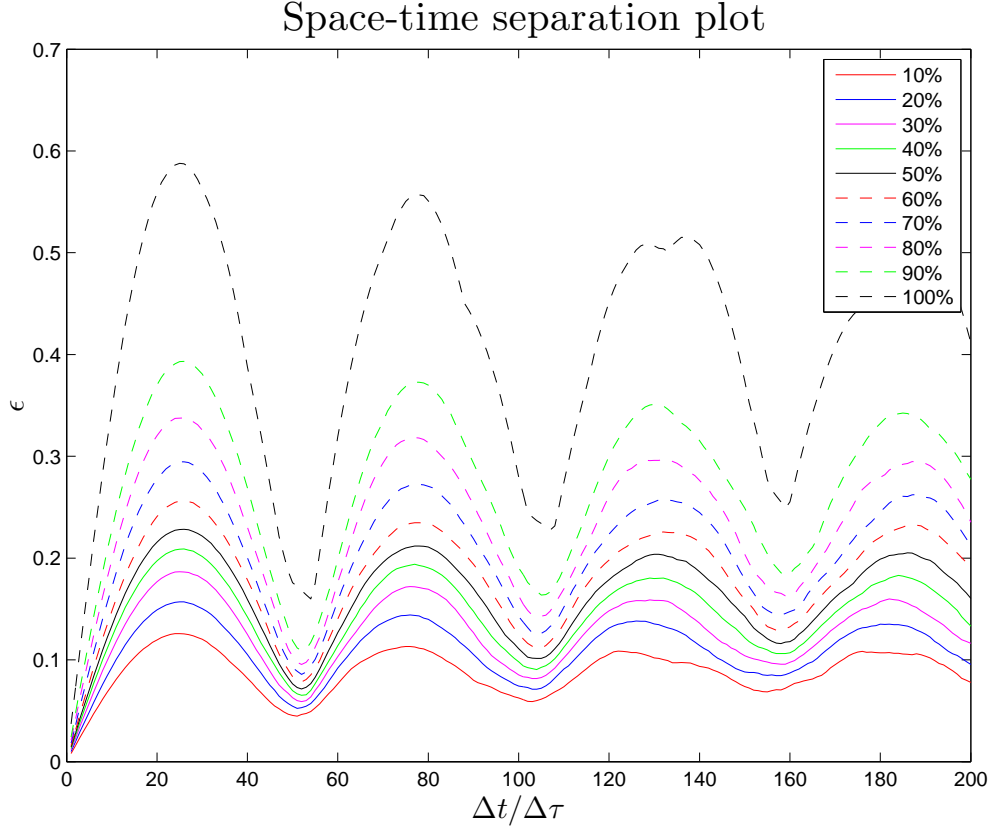}
\caption{\label{fig:SpaceTimeSeparationPlot}The space-time separation plot corresponding to the filtered luminosity measurements of PG 1159-035: $\epsilon$ is the neighbourhood size and $\Delta t$ is the temporal separation 
between two embedding vectors; the optimal embedding dimension $m_0=10$ was used in this plot. The values on the horizontal axis represent time steps in the original time-series arrays ($\Delta \tau=10$ s). The curves 
correspond to different percentiles of the distribution of the distance between two embedding vectors (as detailed in the legend embedded in the plot).}
\vspace{0.35cm}
\end{center}
\end{figure}

The STSP, obtained from the available time-series arrays of PG 1159-035, is displayed in Fig.~\ref{fig:SpaceTimeSeparationPlot}. An oscillatory pattern is seen, reflecting the periodicity of the input signal. (Another 
manifestation of this periodicity is the presence of undulations in the out-of-sample prediction-error arrays $S (k)$.) The STSP of PG 1159-035 is similar to the one obtained by Kantz and Schreiber \cite{kantz1997} from the 
measurements of the flow of a viscous fluid between two coaxial cylinders (Taylor-Couette flow). In my opinion, a reasonable choice of $n_{\rm min}$ would be $2-3$ oscillation periods, i.e., $n_{\rm min} \approx 100-150$ 
(that is, about the optimal embedding window in this study); Kantz and Schreiber would probably recommend the use of an even larger $n_{\rm min}$ value.

\section{\label{sec:Conclusions}Discussion and conclusions}

The goal in this work was the re-analysis of the luminosity measurements of the pre-white dwarf PG 1159-035, in fact those of the observations which found their way into the `Time-series data source Archives: Santa F\'e 
time-series arrays Competition' \cite{timeseries}. Two changes were implemented over the analysis of Ref.~\cite{matsinos2018}:
\begin{itemize}
\item The original time-series arrays were subjected to a symmetrical, two-sided filtering procedure in this paper; on the contrary, the filtering was performed `left-to-right' in Ref.~\cite{matsinos2018}. The comparison 
of the filtered arrays between the two studies had indicated that the impact of this modification on the important results would not be significant.
\item Unlike in Ref.~\cite{matsinos2018}, where the training and test sets were fixed at the outset of the study, the creation of a dynamical database, depending on the choice of the input test file, was employed in the 
present analysis. Given the substantial increase of the database (from Ref.~\cite{matsinos2018} to this work), the significance of this modification needed to be assessed.
\end{itemize}

The results of the analysis of the luminosity measurements of the pre-white dwarf PG 1159-035, obtained in this work, are similar to those reported in Ref.~\cite{matsinos2018}. Evidently, the two aforementioned modifications 
do not affect the important results and the conclusions drawn in Ref.~\cite{matsinos2018}. In particular, the maximal Lyapunov exponent $\lambda$, associated with the variation of the luminosity of PG 1159-035, came out 
equal to $(8.9 \pm 0.7 ({\rm stat.}) \pm 1.9 ({\rm syst.})) \cdot 10^{-2}~\Delta \tau^{-1}$, in very good agreement with the result of Ref.~\cite{matsinos2018}. ($\Delta \tau$ represents the sampling interval in the 
measurements, namely $10$ s).

In relation to the subject of investigation in Ref.~\cite{matsinos2018} and in the present work, I doubt that more information can be extracted from the luminosity measurements of PG 1159-035 found in Ref.~\cite{timeseries}. 
Although there is no indication that an enhanced database will lead to significant changes in the reported results, I will nevertheless express my interest again in receiving the complete data set of the 1989 runs from a 
credible source, e.g., directly from one of the members of the Whole Earth Telescope Collaboration.

\begin{ack}
All the figures were created with MATLAB \textregistered~(The MathWorks, Inc., Natick, Massachusetts, United States).
\end{ack}

\end{document}